\documentclass[12pt,preprint]{aastex}
\newcommand{\hi}{H\,{\sc i}}

\newcommand \kms{km~s$^{-1}$} 
\newcommand{\hii}{H\,{\sc ii}}

\begin{document}

\title{AST/RO Observations of CO $J=4 \rightarrow 3$
 Emission from the N44 Complex in the Large Magellanic Cloud}
\author{Sungeun Kim\altaffilmark{1}, Wilfred Walsh\altaffilmark{2}, \& Kecheng Xiao\altaffilmark{2}}
\altaffiltext{1}{Department of Astronomy \& Space Science, Sejong University, KwangJin-gu, 
KunJa-dong 98, Seoul, 143-747, Korea; e-mail: skim@arcsec.sejong.ac.kr}
\altaffiltext{2}{Harvard-Smithsonian Center for Astrophysics, 60 Garden St., MS-12, Cambrdige, MA 
02138, USA}

\begin{abstract}
We present Antarctic Submillimeter Telescope and Remote Observatory (AST/RO) observations of 
$^{12}$CO $J=4\rightarrow3$ and $^{12}$[C\,I] emission in the N44 H\,II complex in the Large
Magellanic Cloud. We detected strong $^{12}$CO $J=4\rightarrow3$ emission toward the H\,II region 
called as N44BC, which is located on the rim of an expanding giant shell in the N44 region. 
Analysis with a photodissociation region (PDR) model showed that the $^{12}$CO $J=4\rightarrow3$ emitting 
cloud is very dense, with $n_0\approx 10^5$ cm$^{-3}$. We also note that there is a high-velocity component 
associated with the $^{12}$CO $J=4\rightarrow3$ emission. This probably originates from molecular material 
accelerated as a result of the motion induced by the expanding giant shell surrounding LH47 in the N44 
complex. We found that the kinetic energy of this high-velocity gas observed in the CO $J=4\rightarrow3$ 
emission toward the rim of the expanding H\,II shell is at least an order of magnitude higher than the 
kinetic energy derived for the H\,I and H\,II gas in this region. 
\end{abstract}


\section{Introduction}
\label{s:intro}                       

The formation of molecular hydrogen plays a key role in star formation
on all scales. We have learned over the last two decades that the
interactions between atomic clouds and SNRs and/or stellar winds can
provide a mechanism by which molecules can form (Elmegreen 1993). Furthermore, 
observations of atomic hydrogen in our Milky Way and nearby galaxies reveal the 
ubiquity of shells and supershells in the interstellar medium (ISM), indicating 
a potential origin for the substantial molecular material found in the interstellar 
medium.

The Large Magellanic Cloud (LMC) presents us with a unique opportunity to study the
interaction of massive stars and their interstellar environment that perfectly illustrated by 
the Spitzer early release observations of 30 Doradus in the LMC (Brandl et al. 2004). First, 
the stars in the LMC are at a common distance and are close enough that individual stars and 
their stellar ejecta can be studied in great detail. Second, the LMC is inclined at only 27 
degrees to the line of sight, so that the 3-d structure of mass-loss bubbles can be mapped 
without problems of confusion, and third, the reddening is low in virtually all fields. Therefore, 
the LMC provides an excellent opportunity to study the effects of different levels of UV radiation 
from stars on their environments in the multi-phase ISM, and allows us to apply our understanding of
the interstellar properties to studies of the early evolution of high redshift and metal poor galaxies.

At a distance of 55 kpc (Feast 1991), the LMC can be mapped with a high spatial resolution at 
high excitation CO lines by the Antarctic Submillimeter Telescope and Remote Observatory (AST/RO).
A recent [C I] and CO $J=4\rightarrow3$ study of the N159/N160 complexes in the LMC by Bolatto 
et al. (2000) elucidated the condition of atomic and molecular medium in the early stages of star 
formation in this low metallicity system. The dust-to-gas ratios are lower than those in the 
Milky Way, so UV radiation can penetrate deeper into the clouds and dissociate more CO molecules 
to greater depths in the LMC (Pak et al. 1998; Bolatto et al. 1999; Bolatto, Jackson, \& Ingalls 1999). 
The present study aimed to look for dense molecular clouds in the swept-up material of the H\,I 
supershells in the LMC as support for self-propagating star formation (Elmegreen \& Lada 1977; 
Dopita, Ford, \& Webster 1985; Olsen et al. 2001) and found the CO $J=4\rightarrow3$ emission 
toward the rim of the expanding giant H\,I and H\,II shells surrounding the OB association LH 47 
(Lucke \& Hodge 1970; Chu \& Mac Low 1990; Meaburn \& Laspias 1991; Hunter 1994; Oey \& Massey 
1995; Magnier et al. 1996; Kim et al. 1998a). Compared to the CO $J=1\rightarrow0$ emission (Cohen 
et al. 1988; Israel et al. 1993; Chin et al. 1997; Heikkil\"{a} et al. 1999; Fukui et al. 1999; 
Mizuno et al. 2003), this traces the massive core of the molecular clouds. 

\section{Observations}

The observations were performed during the austral winter seasons of 2002 at the Antarctic 
Submillimeter Telescope and Remote Observatory (AST/RO), located at 2847 m altitude at the Amundsen-Scott 
South Pole Station (Stark et al. 2001).  This site has very low water vapor, high atmospheric 
stability and a thin troposphere, making it exceptionally good for submillimeter observations 
(Chamberlin, Lane, \& Stark 1997; Lane 1998). AST/RO is a 1.7-m diameter, offset Gregorian 
telescope that is capable of observing at wavelengths between 200 $\mu$m and 1.3 mm (Stark et al. 2001). 
The receivers used were a 230 GHz SIS receiver with 75-90 K double-sideband noise temperature and 
a dual-channel SIS waveguide receiver (Walker et al. 1992; Honingh et al. 1997) for simultaneous 
461--492 GHz and 807 GHz observations, with double-sideband noise temperatures of 320--390 K and 
1050--1190 K, respectively. Telescope efficiency, $\eta_{\ell}$, estimated using moon scans, skydips, 
and measurements of the beam edge taper, was $\sim$90\% at 230 GHz, 81\% at 461--492 GHz, and 71\% 
at 807 GHz.

Atmosphere-corrected system temperatures ranged from 200 to 400 K at 219-230 GHz, 700 to 4000 K at 
461--492 GHz and 9000 to 75,000 K at 807 GHz.
A beam switching mode was used, with emission-free reference positions chosen at least $20\arcmin$ 
from the regions of interest, to make a small map of points surrounding DEM152. These maps were repeated 
as often as required to achieve a suitable signal--to--noise ratio.

Emission from the CO $J=4\rightarrow3$, and CO $J=7\rightarrow6$ lines at 461.041 GHz and 806.652 GHz, 
together with the [C\,I] line at 492.262 GHz, was imaged over an six arcminute square region centered 
on 5$^{\rm h}$22$^{\rm m}$, $-67^\circ58'$ (J2000) with $0.5\arcmin$ spacing; i.e., a spacing of a 
half-beamwidth or less. AST/RO suffers pointing errors of the order of $1\arcmin$, and the beam sizes 
(FWHM) were $103$--$109\arcsec$ at 461--492 GHz and $58\arcsec$ at 807 GHz (Stark et al. 2001). 
%To facilitate comparison of the various transitions, the data were regridded 
%onto a $0.25\arcmin$ grid and smoothed to a FWHM spatial resolution of $3\arcmin$ with a Gaussian 
%filter function.

Two acousto-optical spectrometers (Schieder, Tolls, and Winnewisser 1989) were used as backends. 
The AOSs had 1.07~MHz resolution and 0.75 GHz effective bandwidth, resulting in velocity resolution 
of 0.65 \kms\ at 461 GHz and 0.37 \kms\ at 807 GHz.  
The data were then smoothed to a uniform velocity resolution of 1 \kms\ .  The high frequency 
observations were made with the CO $J = 7 \rightarrow 6$ line in the lower sideband (LSB). Since 
the intermediate frequency of the AST/RO system is 1.5 GHz, the $^3P_2 \rightarrow ^3 \! P_1$ line 
of [C\,I] line appears in the upper sideband (USB) and is superposed on the observed LSB spectrum. 
The local oscillator frequency was chosen so that the nominal line centers appear separated by 100 
\kms\ in the double-sideband spectra. A third AOS, used for only a few spectra, had 0.031 MHz 
resolution and 0.25 GHz bandwidth.

The standard chopper wheel calibration technique was employed,
implemented at AST/RO by way of regular (every few minutes)
observations of the sky and two blackbody loads of known temperature (Stark et al. 2001). 
Atmospheric transmission was monitored by regular skydips, and known, bright sources were observed 
every few hours to further check calibration and pointing. At periodic intervals and after tuning, 
the receivers were manually calibrated against a liquid-nitrogen-temperature load and the two blackbody 
loads at ambient temperature and about 100 K. The latter process also corrects for the dark current of 
the AOS optical CCDs. The intensity calibration errors became as large as $\pm15$\% during poor weather 
periods.

The data in this survey were reduced using the COMB data reduction package.  After elimination of 
scans deemed faulty for various instrumental or weather-related reasons (less than $\sim 10\%$ of 
the total dataset), linear baselines were removed from the spectra.

\section{RESULTS}
\label{s:results}

In Figure 1, we present a contour map of the CO $J=4\rightarrow3$ line emission taken with 
AST/RO overlaid on an H$\alpha$ image. There is strong CO $J=4\rightarrow3$ emission 
detected at RA=5$^{\rm h}$22$^{\rm m}$05$^{\rm s}.3$, Dec=$-$67$^\circ$58$'$41$''$.8 (J2000) 
which is associated with N44 complex in the LMC. The interstellar complex N44 was found 
by Henize in 1956 and is one of the largest complexes in the LMC. It consists of two giant 
H$\alpha$ shells. 'Shell 1' is prominent; the other one is fainter and located at the western 
side of 'Shell 1' (Meaburn \& Laspias 1991). 
'Shell 1' contains the OB association LH 47 (Lucke \& Hodge 1970). The CO $J=4\rightarrow3$ emitting 
cloud is located along the western boundary of the giant \hii\ shell 'Shell 1' emission, as seen in 
Figure 1, and the peak of the CO $J=4\rightarrow3$ emission appears to be associated 
with the bright \hii\ regions N44 BC ($\alpha_{1950}=5^{\rm h}22^{\rm m}10^{\rm s}.6$, 
$\delta_{1950}=-68^{\circ}00'32''$). The N44 complex is one of the brightest CO $J=1\rightarrow0$ 
emission regions in the LMC from the 1.2-m MINI telescope survey by Cohen et al. (1988) and the 
recent 4-m NANTEN survey by Fukui et al. (1999) and Mizuno et al. (2003). DCO$^{+}$ emission has 
also been detected in N44 BC (Chin et al. 1996). 

Figure 2 presents the 461 GHz CO $J=4\rightarrow3$, 492 GHz [C\,I], 230 GHz CO $J=2\rightarrow1$ line 
profiles at the peak of the contour map at RA=5$^{\rm h}$22$^{\rm m}$5$^{\rm s}.3$, 
Dec=$-$67$^\circ$58$'$41$''$.8 (J2000). The 461 GHz CO $J=4\rightarrow3$ emission from this region
is bright, with $T_{MB}$=2.1 K. The peak of the CO $J=4\rightarrow3$ emission appears at 
$V_{LSR}$=283.1 \kms\ as seen in Figure 2, which is similar to the \hi\ systemic velocity of 281 
\kms\ at this position (Kim et al. 1998a). 
The FHWM of CO $J=4\rightarrow3$ emission is 7.5 $\pm 0.1$ \kms\ . This is somewhat narrower than 
the FWHM of \hi\ emission, which is about 9 \kms\ . Figure 3 shows the average spectrum for the entire
AST/RO observations of the 461 GHz CO $J=4\rightarrow3$ emission and compares it to that of the 492 GHz
[C\,I] emission line. We note a high-velocity component of CO $J=4\rightarrow3$ emitting molecular gas 
at $V_{LSR} \approx 304$ \kms\ as seen in this Figure. The average [C\,I] 492 GHz line profile over the 
entire AST/RO observation gives a much lower temperature for the [C\,I] line, $T_{MB}$=0.13 K at 
$V_{LSR}$=283.1 \kms\ . The line parameters for these spectra are given in Table 1.
The luminosity $L_{CO}$, arising from the $^{12}$CO $J=4\rightarrow3$ emission is about 3.73 
$\times$ 10$^{3}$ $L_\odot$, calculated as the product of the area and the integral of the mean 
global profile. 

\section{DISCUSSION}
\subsection{Photodissociated Regions in N44 Complex}

Hydrogen is the most abundant element in the universe, and hydrogen atoms also make up most of the
interstellar matter found in the LMC (Luks \& Rolhfs 1992; Kim et al. 2003). A comparison of the 
CO $J=4\rightarrow3$ emission and \hi\ emission from the N44 complex clearly shows that the 
position of peak brightness temperature in $^{12}$CO $J=4\rightarrow3$ emission corresponds 
to the region where the \hi\ emission is deficient, as seen in Figure 4. The \hi\ image was made from 
an aperture synthesis survey of the LMC with the Australia Telescope Compact Array (ATCA) (Kim et al. 
1998b) and the Parkes single-dish survey (Staveley-Smith et al. 2003). The two surveys were combined 
using a Fourier-plane technique (Kim et al. 2003).

It is not yet possible to determine the CO $J=4\rightarrow3$ emission concentrations within the 
whole N44 complex in the LMC with our limited spatial coverage. However, our results show the  
detection of massive molecular clouds associated with the giant H$\alpha$ shell within the N44 
complex in the LMC. The observed $^{12}$CO $J=4\rightarrow3$ emission indicates that the neutral 
ISM toward the rim of the giant \hii\ shell surrounding LH47 in the N44 complex is very dense and 
warm, since the $^{12}$CO $J=4\rightarrow3$ transition requires $T>50K$ and $n\sim10^5$ cm$^{-3}$
(Bolatto et al. 2000; Cecchi-Pestellini et al. 2001; Zhang et al. 2001).  
%Though the CO brightness is more an indicator of temperature than that of the total amount of 
%molecular gas (Allen \& Lequeux 1993; Willacy, Langer, \& Allen 2002; Yamamoto et al. 2003).

%CHECK NUMBERS!!!
The column density of neutral hydrogen, $N$(H\,I)=$1.822 \times 10^{18} \int T_b~dv$ (Spitzer 1978), at 
the peak of the CO $J=4\rightarrow3$ emission was derived to be $\sim$3.4 $\pm0.1$ $\times$ 10$^{21}$ 
cm$^{-2}$. 
By assuming that N44 giant shell 'Shell 1' is situated in the mid-plane of the LMC, we derived the 
volume density of the neutral hydrogen associated with the CO $J=4\rightarrow3$ emitting molecular 
cloud using the scale-height of the LMC which is approximately 180 pc (Kim et al. 1999). We found a 
volume density of neutral hydrogen associated with the CO $J=4\rightarrow3$ emission of $n_H$ $\approx$ 
6.3 $\pm0.1$ cm$^{-3}$. 

For comparison, we attempted to derive the density of hydrogen molecules, $n_{H_2}$, in the $^{12}$CO 
$J=4\rightarrow3$ emitting gas in the N44 complex. As we have discussed above, the $^{12}$CO 
$J=4\rightarrow3$ emission traces the warm and dense molecular gas where collisional de-excitation 
balances radiative de-excitation (Spitzer 1978; Hollenbach \& Tielens 1999). This usually occurs 
when the far-ultraviolet (FUV) photons from stars photodissociate the molecular cloud (Tielens \& 
Hollenbach 1985; Sternberg \& Dalgarno 1989; Wolfire, Tielens, \& Hollenbach 1990; Hollenbach \& Tielens 
1990). 
The $^{12}$CO $J=4\rightarrow3$ emitting molecular cloud associated with the N44 complex is a  
fine example of a photodissociated region (PDR) in the LMC. The FUV photons (6 eV  $<$ $h\nu$ $<$ 
13.6 eV) of stars in the OB association, LH 47, enclosed by Shell 1 and a single ionizing star within 
N44 C (Stasinska et al. 1986; Meaburn \& Laspias 1991) are likely to be responsible for the 
photodissociation. 

Using our newly observed line ratios of $^{12}$CO $J=4\rightarrow3$ and [C\,I] emission, the PDR 
model allows us to estimate the cloud density, kinetic temperature, and the FUV fluxes of the gas 
(Tielens \& Hollenbach 1985; Sternberg \& Dalgarno 1989; Wolfire, Tielens, \& Hollenbach 1990; 
Sternberg \& Dalgarno 1995; Kaufman et al. 1999). This is illustrated in Figure 5, where the observed
line ratios of the CO $J =2 \rightarrow 1$, $J =4 \rightarrow3$, and [C\,I] 609 $\mu$m emission lines are 
plotted against an estimate of the cloud density $n$ and an estimate of the incident FUV flux $G_0$. 
$G_0$ is in units of the local interstellar value of 1.6 $\times10^{-3}$ ergs cm$^{-2}$s$^{-1}$ 
(Habing Field). The required particle density of $H_2$ is somewhat greater than $10^5$ cm$^{-3}$.
As the number density of the atomic hydrogen is only a few cm$^{-3}$, the estimated cloud 
density $n$ from the PDR analysis indicates the density of hydrogen molecules in the cloud. We therefore 
conclude that the PDRs in the N44 complex contain about 1.7$\times$10$^5$ hydrogen nuclei per cm$^{-3}$.

This is in agreement with the results of Heikkila et al. (1999) who found a cloud density of 
$n\approx5\times10^5$ cm$^{-3}$ toward the main molecular peaks in the region towards N44 BC by 
multi-line excitation analysis using the $^{13}$CO $J = 1\rightarrow0$/$^{18}$CO $J = 1\rightarrow0$, 
$^{13}$CO $J = 1\rightarrow0$/$^{17}$CO $J = 1\rightarrow0$, and $^{18}$CO $J = 1\rightarrow0$/$^{17}$CO 
$J = 1\rightarrow0$, and $^{18}$CO $J = 1\rightarrow0$/$^{17}$CO $J = 1\rightarrow0$. It is not surprising 
to find such a high density of hydrogen molecules in the region associated with N44 BC, where we detect the 
strong $^{12}$CO $J=4\rightarrow3$ emission. However, it is interesting to confirm the existence of very 
high density concentrations of molecular gas in the expanding H\,I and H$\alpha$ shell.
In addition, the present study has shown the extent of the dense gas as well as its location. A previous 
study by Chin et al. (1997) derived the molecular mass of N44 BC where we detect strong CO $J=4\rightarrow3$
emission, to be about 10$^5$ $M_\odot$ using the virial theorem. If we estimate the molecular mass in the 
shell simply by $dM=2n_{H_2}m_HdV$ and by adopting a deprojected size of the CO $J=4\rightarrow3$ emitting 
cloud, $r_c$, of about 22.7 pc, then we find that the mass of molecular gas associated with N44 BC may be 
at least an order of magnitude greater than the mass derived by the virial theorem.  

\subsection{CO $J=4\rightarrow3$ high-velocity gas?}

As clearly seen in Figure 3, there is a high-velocity component of CO $J=4\rightarrow3$ emission 
at $V_{LSR}$=304 \kms\ arising from the molecular cloud which is associated with expanding H\,I and 
H$\alpha$ shells (Meaburn \& Laspias 1991; Magnier et al. 1996; Kim et al. 1998a).
The brightness temperature of the high-velocity gas, $T_{MB}$ is about 0.1 K. This faint emission is 
probably the receding side of accelerated material with an expansion velocity of $v_{exp}\approx20.9$ 
\kms\ . In this scenario, the high-velocity gas is presumed to be caused by the interaction of the 
molecular cloud with the stellar winds and/or supernovae blasts of the hot stars in LH 47 (Lucke \& Hodge 
1970; Oey \& Massey 1995). The expansion pattern of the H\,I gas shell surrounding LH 47 is similar to 
that of the ionized gas shell of Shell 1 (Kim et al. 1998a). 
The present study shows that the acceleration of the CO $J = 4\rightarrow3$ emitting gas 
is only slightly smaller than the expansion velocity of the \hi\ gas, which is about 30 \kms\ .
We therefore expect that this high-velocity gas in the CO $J = 4\rightarrow3$ emission is 
associated with the expanding atomic and ionized gas shells in the N44 complex. However, it is also 
conceivable that the H\,II region, N44 BC, has driven a shock into the molecular cloud and produced 
such high-velocity molecular gas.
 
We estimate an upper limit on the kinetic energy of the high-velocity molecular cloud of $\sim2.6
\times 10^{51}$ ergs. Assuming that this high-velocity molecular cloud is produced by an outward 
transfer of momentum from the expanding H\,I and H\,II shells, we calculate an upper limit on the 
total kinetic energy of the shell of 2.9 $\times$ $10^{51}$ ergs. We have summed the kinetic energy 
in the ionized gas shell and in the \hi\ shell for the superbubble Shell~1 in N44 (Kim et al. 1998a), 
and the upper limit of the kinetic energy of the high-velocity molecular cloud in N44. Adopting the
thermal energy of Shell 1 (Chu \& Mac Low 1990; Magnier et al. 1996; Kim et al. 1998a), we find that
the sum of the total kinetic energy of the shell and its thermal energy is similar to the energy 
released by stellar winds and supernovae explosions using Weaver et al.'s (1977) pressure-driven 
bubble model. 

\acknowledgments
\label{s:ack}
We thank Antony A. Stark (AST/RO P.I.), Adair P. Lane (AST/RO Project
Manager), and C. Martin (2001 winterover) at SAO; C. Walker and his SORAL receiver group at the
U. of Arizona; J. Kooi and R. Chamberlin of Caltech, G. Wright of
PacketStorm Communications, and K. Jacobs of  U. K\"{o}ln for their work on the instrumentation; R. 
Schieder, J. Stutzki, and colleagues at U. K\"{o}ln for their AOSs. We thank Marc Pound for helping 
us to use PDRT. We thank C. Smith for his H$\alpha$ image and You-Hua Chu for helpful discussion. We 
thank HI project team members: Lister Staveley-Smith, Robert J. Sault, Mike Dopita, Ken C. Freeman, 
Dave McConnell, Mike Kesteven. This research was in part supported by the National Science Foundation 
under a cooperative agreement with the Center for Astrophysical Research in Antarctica (CARA), grant 
number NSF OPP 89-20223. CARA is a National Science Foundation Science and Technology Center. 
Support was also provided by NSF grant number OPP-0126090. SK was in part supported by Korea Science \& 
Engineering Foundation (KOSEF) under a cooperative agreement with the Astrophysical Research Center of 
the Structure and Evolution of the Cosmos (ARCSEC).

%Mizuno, CO NANTEN survey, 282.5 km/sec, Delta V=7.2 km/s, T_R*~1.3 K, 10.1 K km/s;  

\newpage
\begin{table*}
\caption{Observed Line Parameters}
\begin{tabular}{lllll}
\noalign{\smallskip}
\hline\hline
\noalign{\smallskip}
Line                         & $T_{MB}$   & $V_{LSR}$ & $\Delta V$ & $\int T_{MB}$ dV\\
                             &~(K)      &~(\kms\ ) & (\kms\ ) & (K \kms\ ) \\
\noalign{\smallskip}
\hline
461 GHz CO $J=4\rightarrow3$ &  0.443$\pm$ 0.004 & 283.165$\pm$0.036 & 7.391$\pm$0.082 & 3.484\\
492 GHz CI                   & 0.126 $\pm$ 0.027 & 281.683$\pm$0.327 & 3.130$\pm$0.771 & 0.420 \\ 
\noalign{\smallskip}
\hline
\end{tabular}
\end{table*}

\newpage
\begin{figure}
\figurenum{1}
\plotone{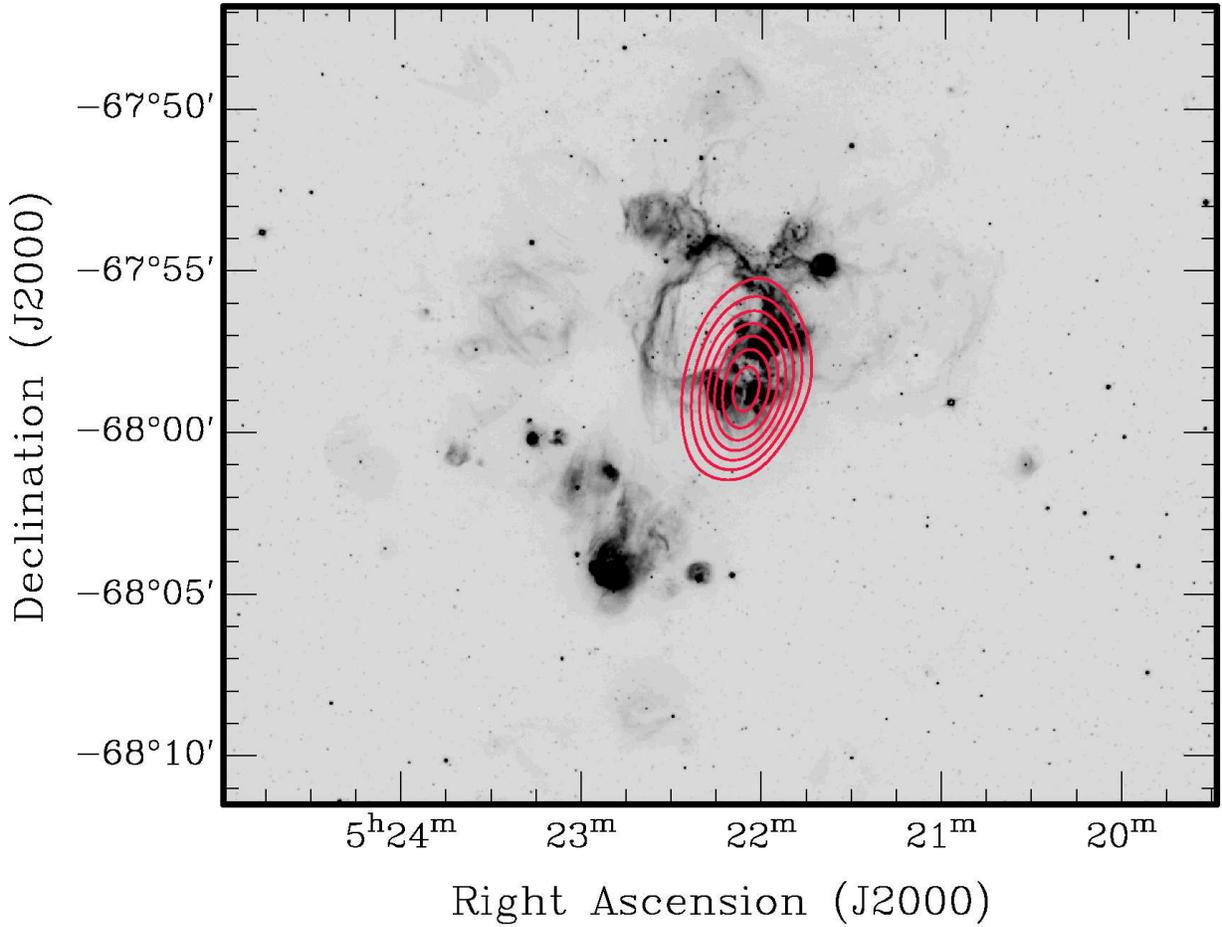}
\caption{Integrated intensity contour map of $^{12}$CO $J= 4 \rightarrow3$ emission observed with the 
AST/RO telescope. The grey scale is the H$\alpha$ image of the N44 complex taken by Smith et al. (1998). 
The range of velocity integration is from 272 \kms\ to 300 \kms\ . The lowest contour is 2.3 K \kms\ and 
the contours increase by 1.0 K \kms\ .}
\end{figure}

\begin{figure}
\figurenum{2}
\plotone{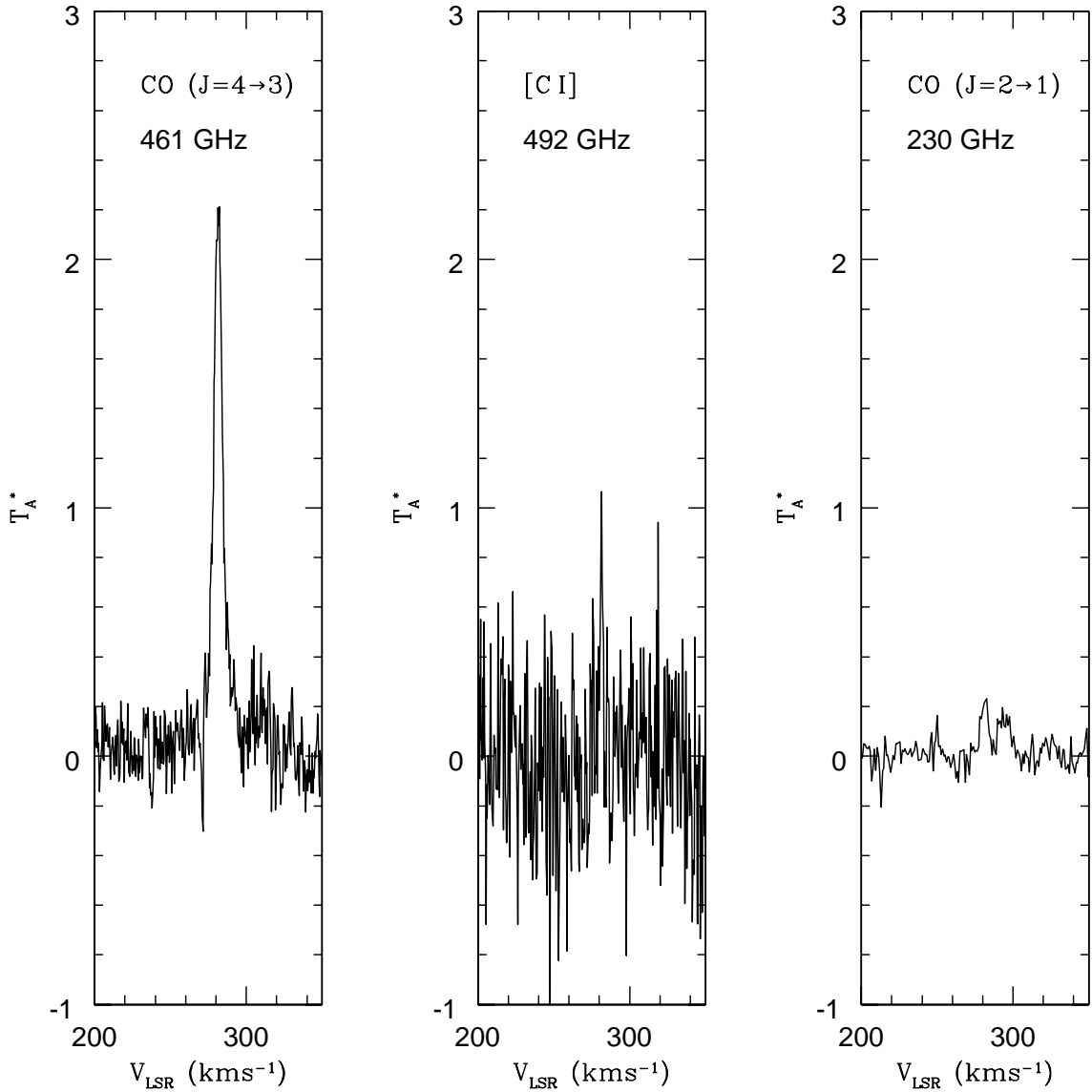}
\caption{The $^{12}$CO $J=4\rightarrow3$, $^{12}$[C\,I] 492 GHz, and $^{12}$CO $J=2\rightarrow1$ spectra 
observed toward the peak of the $^{12}$CO $J=4\rightarrow3$ emission in the N44 complex. A $^{12}$CO 
$J=2\rightarrow1$ line profile was obtained for this position.}
\end{figure}

\begin{figure}
\figurenum{3}
\plotone{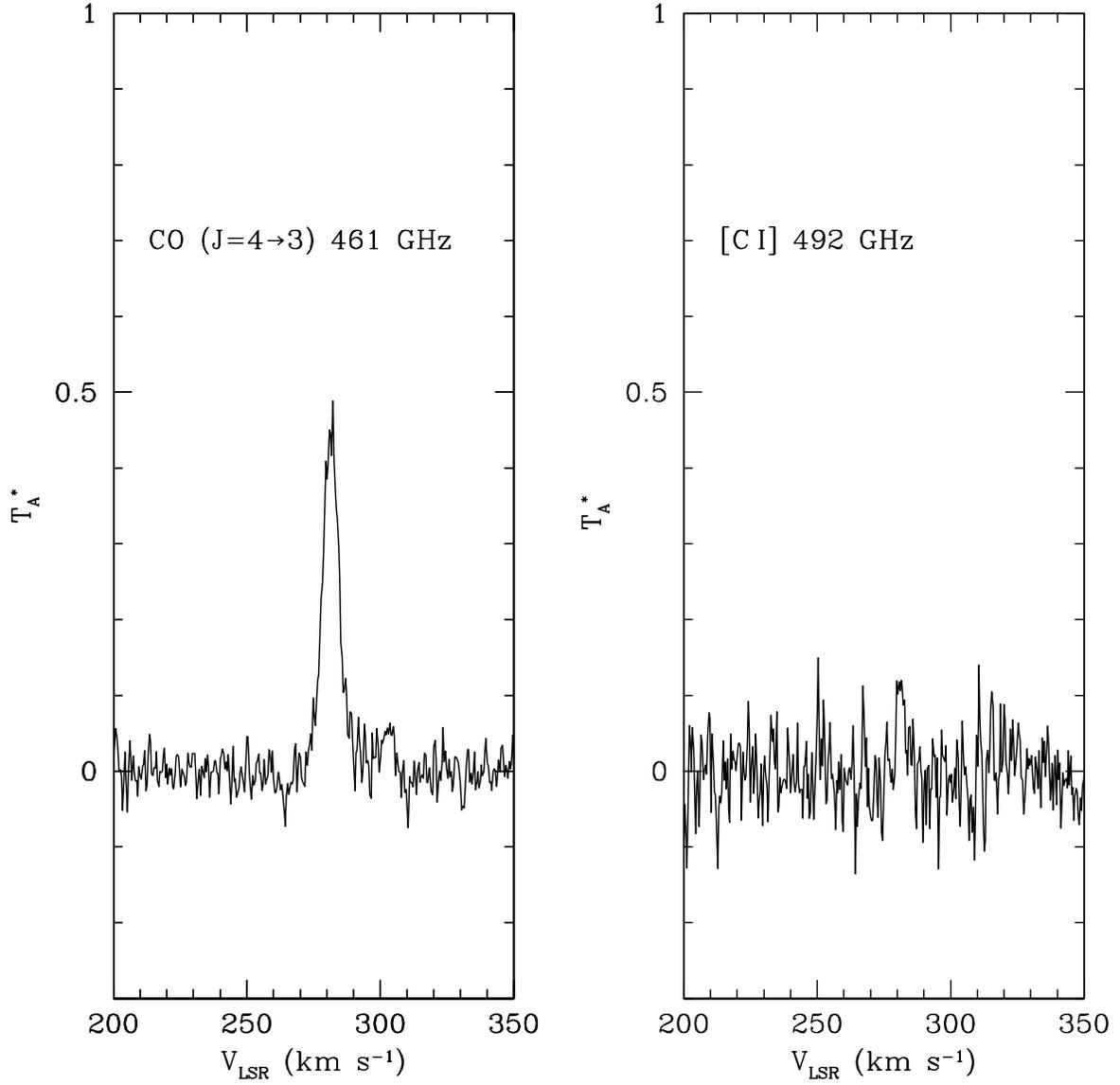}
\caption{The $^{12}$CO $J=4\rightarrow3$ and $^{12}$[C\,I] 492 GHz spectra averaged
over the detected $^{12}$CO $J=4\rightarrow3$ emission in the N44 complex.}
\end{figure}

\begin{figure}
\figurenum{4}
\plotone{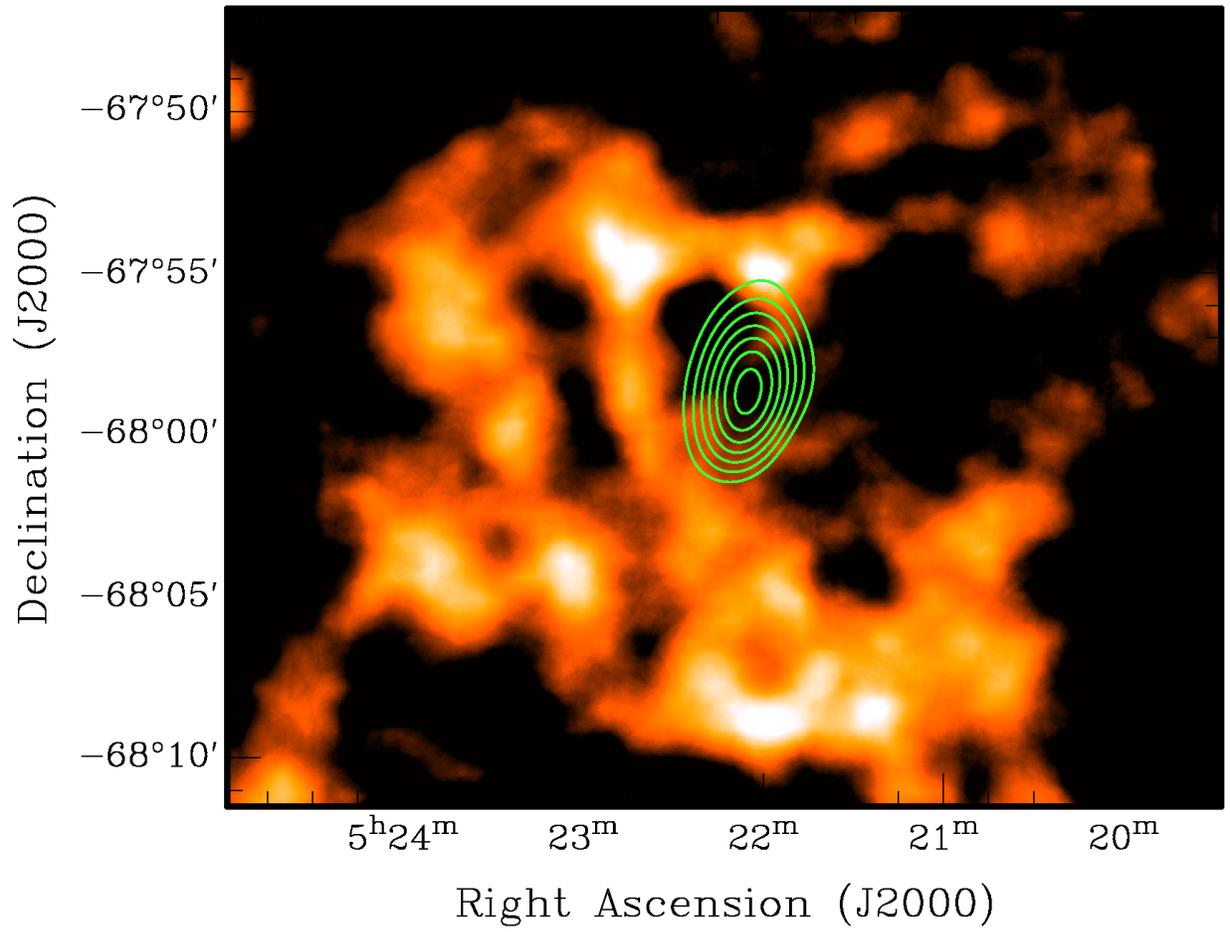}
\caption{The ATCA integrated \hi\ image overlaid with $^{12}$CO $J= 4 \rightarrow3$ emission line contours 
as described in Figure 1.}
\end{figure}

\begin{figure}
\figurenum{5}
\plotone{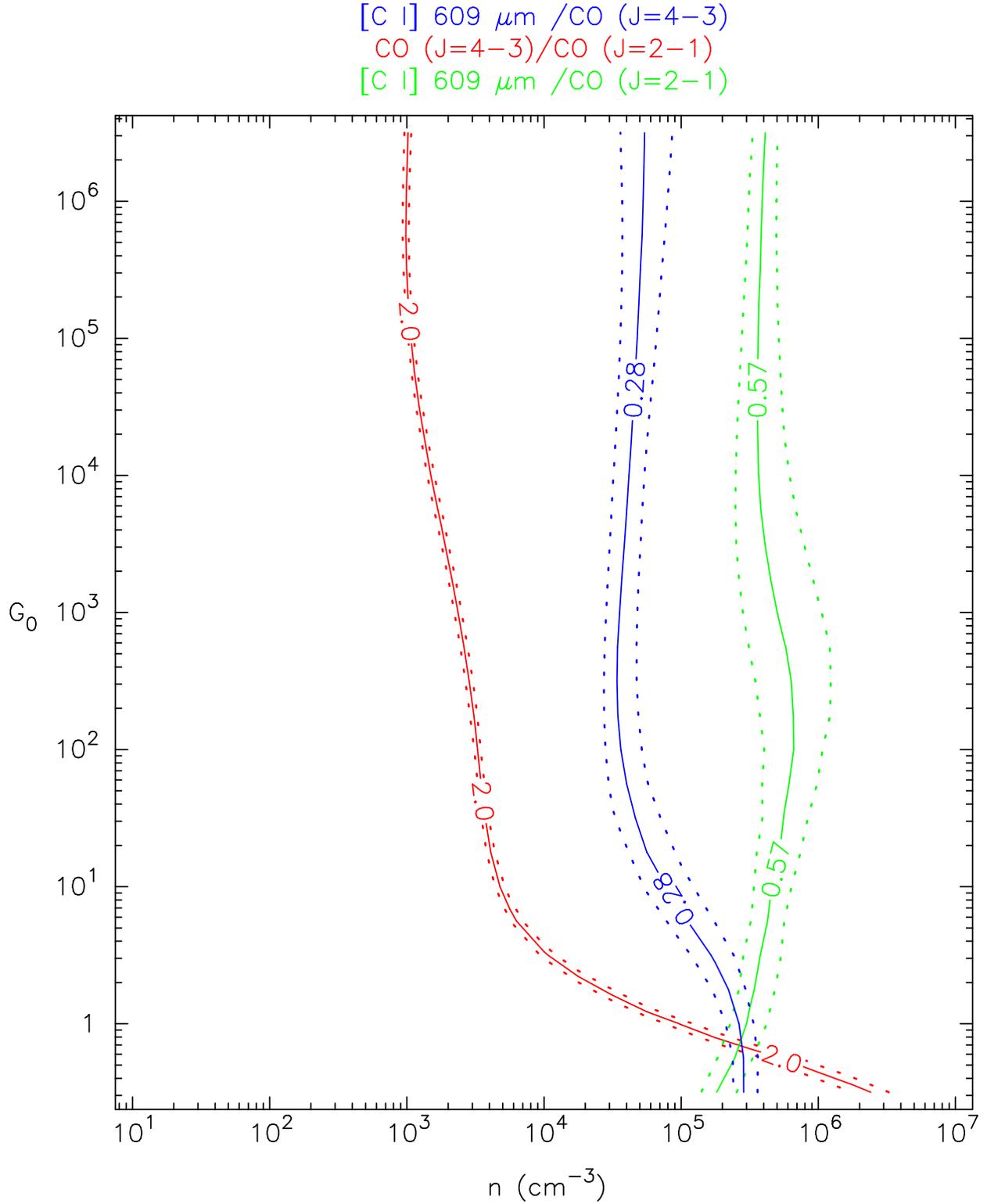}
\caption{Line ratios of the $^{12}$CO $J=4\rightarrow3$, $^{12}$[C\,I] 492 GHz, and $^{12}$CO 
$J=2\rightarrow1$ spectra observed with AST/RO are plotted against an estimate of the cloud 
$n$ density and incident FUV flux $G_0$ obtained using the PDR model.}
\end{figure}

\end{document}